\documentclass[12pt,,letterpaper]{JHEP3}
\usepackage{graphics}
\usepackage{epsfig}
\usepackage{amssymb}
\usepackage{stmaryrd}
\usepackage{amsmath}
\usepackage{amsfonts}
\usepackage{mathrsfs}
\usepackage{graphicx}

\def\be{\begin{equation}}
\def\ee{\end{equation}}
\def\ba{\begin{eqnarray}}
\def\ea{\end{eqnarray}}
\def\nn{\nonumber}

\newcommand{\vp}{\varphi} 

\title{One loop partition function from normal modes for $ \mathcal{N}=1$ supergravity in AdS$_3$}

\author{Hongbao Zhang\\
Crete Center for Theoretical Physics, Department of Physics, \\
University of Crete, 71003 Heraklion, Greece\\
 \email{hzhang@physics.uoc.gr}}

\author{Xiangdong Zhang\\
Department of Physics, Beijing Normal University, 100875 Beijing,
China \\
\email{zhangxiangdong@mail.bnu.edu.cn}}
\preprint{CCTP-2012-11}
\abstract{With the recently discovered formula, which relates the off shell Euclidean one loop determinant to the on shell quantities such as normal and quasinormal modes in real spacetime, we work out the one loop partition function for $\mathcal{N}=1$ supergravity in AdS$_3$ from scratch. In passing, we also provide the explicit expression for one loop determinant of a field with arbitrary spin in AdS$_3$. To achieve this, we firstly derive the determinant expression for the one loop partition function in question using the powerful decomposition technique, and then we construct the normal modes in a purely algebraic way by demonstrating that the space of normal modes falls into the representation of SL(2,$R$) Lie algebra associated to AdS$_3$. The whole procedure developed here turns out to be much simpler than the previous strategy.}

\begin{document}
\section{Introduction}
AdS/CFT correspondence, as an explicit implementation of holographic
principle, relates a $d+1$ dimensional gravitational theory to a $d$
dimensional quantum field theory, where the black hole in the bulk
is dual to the boundary field theory at a finite temperature. Most
of checks and applications of this duality have restricted attention
to the large $N$ limit of boundary field theory, where gravity is
treated as a classical object in the bulk. In order to capture a
$\frac{1}{N}$ correction to the boundary field theory by holography, one is
required to calculate the one loop partition function for a quantum
field propagating around the classical gravitational background
mentioned above in the bulk.

Although it is technically challenging to compute the one loop
partition function in a nontrivial gravitational background, there
have been various methods developed to approximate it if the one
loop partition function can not be determined explicitly. Among others, with the heat kernel method, the one loop partition can be calculated out explicitly for pure gravity in AdS$_3$\cite{GMY}. Later on, such a heat kernel method 
is simplified by using the group theoretical approach and the explicit expression is obtained for the one loop determinant of a field with arbitrary spin in AdS$_3$\cite{DGG}.
This is supposed to be not too surprising from both
bulk and boundary perspectives in the context of AdS$_3$/CFT$_2$ as
there are sufficient symmetries on both sides to guarantee the
integrability.

On the other hand, recently the authors in \cite{DHS} have derived a new
formula to evaluate the one loop bulk partition function in terms of normal or quasinormal modes, which arise as the
solutions to the equation of motion for the field in question with
the appropriate boundary conditions. This result is highly impressive
because as alluded to above it relates the off shell object to the
on shell quantities somehow. On the other hand, in the context of
AdS/CFT correspondence the concept of normal and quasinormal modes is of
interest in its own right. These modes correpond to the excitations on the boundary, as by the holographic recipe they
arise as the poles of the retarded Green function of the dual
operator.

Now with the above new formula for the one loop partition function,
the task boils essentially down to the determination of normal or quasinormal
modes, depending on whether the bulk has horizon or not. The approximate expressions for these modes offer a
method for one to approximate the one loop partition function. In
particular, the one loop partition function can be exactly evaluated
through this new formula in those cases where the complete spectrum
of normal or quasinormal modes can be nailed down. One of such cases is 
AdS$_3$.  Actually as shown in \cite{MS} and \cite{BKL}, the spectrum of normal modes
can be obtained as an infinite tower of descendants of the highest weight mode associated with the SL(2,$R$) Lie
algebra of Killing fields for a scalar field in AdS$_3$. 

The purpose of this paper is two fold. One is to construct the spectrum of normal modes in a similar way for any other higher spin field in AdS$_3$ by demonstrating that the space of solutions to the equation of motion for the field in question falls into the representation of SL(2,$R$) Lie algebra. To achieve this, we shall employ the simplified computational strategy by sticking to the covariant derivative and introducing
a pair of intrinsic tensor fields associated with the SL(2,$R$)
symmetry\cite{Zhang}. The other is to demonstrate the advantage of the above new formula by working out the one loop partition function from normal modes for $\mathcal{N}=1$ supergravity in AdS$_3$, where instead of resorting to the Faddeev-Popov trick we shall derive the determinant expression for the one loop partition function in question by the decomposition technique, which turns out to be extremely efficient.

The rest of paper is organized as follows.  In the next section we shall
provide a brief review of the SL(2,$R$) Lie algebra of Killing
fields in AdS$_3$ as well as the two types of  intrinsic tensor
fields associated with the SL(2,$R$) quadratic Casimir operator.
Then we describe the dynamics of $\mathcal{N}=1$ supergravity in AdS$_3$ in Section $3$, where in particular we show how the corresponding
one loop partition function can be worked out in terms of Laplace
like determinants by the powerful decomposition technique. In Section 4, we will demonstrate how the
spectrum of normal modes can be constructed in the algebraic manner as
simple as possible by working explicitly on the case of gravitino field. In the
subsequent section, as an application of the one loop determinant expression derived from normal modes for a field with arbitrary spin in AdS$_3$, we evaluate the one loop
partition function for $\mathcal{N}=1$ supergravity, which turns out to be exactly the same as that obtained by the heat
kernal method. We end up with some discussions in the final section and relegate various additional details to the four appendices.

Notation and conventions follow \cite{Wald} unless specified
otherwise.

\section{AdS$_3$ and its SL(2,$R$) symmetries}
Start from the metric for AdS$_3$, i.e.,
\begin{equation}
ds^2=-\cosh^2(\rho)dt^2+\sinh^2(\rho)d\phi^2+d\rho^2,
\end{equation}
where $0\leq\rho\leq\infty$,
$-\infty\leq t\leq\infty$, and $-\pi\leq\phi\leq\pi$. Whence the corresponding curvature reads
\begin{equation}
R_{abcd}=g_{ad}g_{bc}-g_{ac}g_{bd}, R_{ab}=-2g_{ab}, R=-6,
\end{equation}
whereby $AdS_3$ is a space of constant curvature, thus admits
six Killing fields. We denote these six Killing fields by
$L_k$ and $\bar{L}_k$ with $k=0,\pm1$. In particular, $L_k$ is given
by
\begin{eqnarray}
L_0^a&=&i(\frac{\partial}{\partial u})^a, \nonumber\\
L_{-1}^a&=&ie^{-iu}[\frac{\cosh(2\rho)}{\sinh(2\rho)}(\frac{\partial}{\partial
u})^a-\frac{1}{\sinh(2\rho)}(\frac{\partial}{\partial
v})^a+\frac{i}{2}(\frac{\partial}{\partial\rho})^a],\nonumber\\
L_{+1}^a&=&ie^{iu}[\frac{\cosh(2\rho)}{\sinh(2\rho)}(\frac{\partial}{\partial
u})^a-\frac{1}{\sinh(2\rho)}(\frac{\partial}{\partial
v})^a-\frac{i}{2}(\frac{\partial}{\partial\rho})^a],\label{Killing}
\end{eqnarray}
where we have employed another coordinate system, namely
$\{\rho, u=t+\phi,v=t-\phi\}$. $\bar{L}_k$ can be
similarly defined as (2.6) except switch $u$ and $v$ therein.
 Then it can be shown that their Lie commutators satisfy
two sets of the SL(2,$R$) Lie algebra, i.e.,
\begin{equation}\label{commutation}
[L_0,L_{\pm1}]=\mp L_{\pm1},[L_{+1},L_{-1}]=2L_0,
[\bar{L}_0,\bar{L}_{\pm1}]=\mp
\bar{L}_{\pm1},[\bar{L}_{+1},\bar{L}_{-1}]=2\bar{L}_0, [L_k,\bar{L}_l]=0.
\end{equation}
Note that the Lie derivative observes
$[\mathcal{L}_X,\mathcal{L}_Y]=\mathcal{L}_{[X,Y]}$ and
$\mathcal{L}_{\alpha X}=\alpha\mathcal{L}_{X}$ for the arbitrary Killing
vector fields $X$ and $Y$ with the arbitrary constant
$\alpha$. Thus
the above Lie algebra can be naturally represented by the Lie
derivative. In particular, the quadratic Casimir operators can be
realized by the Lie derivative as
\begin{equation}
\mathcal{L}^2=\mathcal{L}_{L_0}\mathcal{L}_{L_0}-\frac{1}{2}(\mathcal{L}_{L_{+1}}\mathcal{L}_{L_{-1}}+\mathcal{L}_{L_{-1}}\mathcal{L}_{L_{+1}}),
\bar{\mathcal{L}}^2=\mathcal{L}_{\bar{L}_0}\mathcal{L}_{\bar{L}_0}-\frac{1}{2}(\mathcal{L}_{\bar{L}_{+1}}\mathcal{L}_{\bar{L}_{-1}}+\mathcal{L}_{\bar{L}_{-1}}\mathcal{L}_{\bar{L}_{+1}}),\label{commutation}
\end{equation}
which commute with both $\mathcal{L}_{L_k}$ and
$\mathcal{L}_{\bar{L}_k}$.

As advocated in \cite{Zhang}, it turns out to be convenient for tensor and spinor analysis related to the above quadratic SL(2,$R$) Casimir  if one constructs the two types of auxiliary tensor fields as follows\footnote{Although the whole construction is built up for the BTZ black hole in \cite{Zhang}, it can be readily translated into AdS$_3$ because AdS$_3$ is related to the BTZ black hole in \cite{Zhang} by change of coordinates as $\{t\rightarrow i\phi,\phi\rightarrow it, \rho\rightarrow\rho\}$.}

\begin{equation}
H^{ab}=L_0^aL_0^b-\frac{1}{2}(L_{+1}^aL_{-1}^b+L_{-1}^aL_{+1}^b),\bar{H}^{ab}=\bar{L}_0^a\bar{L}_0^b-\frac{1}{2}(\bar{L}_{+1}^a\bar{L}_{-1}^b+\bar{L}_{-1}^a\bar{L}_{+1}^b),
\end{equation}
and
\begin{eqnarray}
Z_{abc}&=&L_{0a}\nabla_bL_{0c}-\frac{1}{2}(L_{+1a}\nabla_bL_{-1c}+L_{-1a}\nabla_bL_{+1c}),\nonumber\\
\bar{Z}_{abc}&=&\bar{L}_{0a}\nabla_b\bar{L}_{0c}-\frac{1}{2}(\bar{L}_{+1a}\nabla_b\bar{L}_{-1c}+\bar{L}_{-1a}\nabla_b\bar{L}_{+1c}).
\end{eqnarray}
The essential reason is that they possess the following nice properties, i.e.,
\begin{equation}
H^{ab}=\bar{H}^{ab}=\frac{1}{4}g^{ab},
\end{equation}
and
\begin{equation}
Z_{abc}=\frac{1}{4}\epsilon_{abc},\bar{Z}_{abc}=-\frac{1}{4}\epsilon_{abc},
\end{equation}
where $g$ and $\epsilon$ are the metric and volume element of our AdS$_3$ respectively.
\section{$\mathcal{N}=1$ supergravity and its determinant expression for one loop partition function}
\subsection{Graviton}
Let us start to consider the dynamics of graviton $h_{ab}$ on top of AdS$_3$ by focusing on the perturbative expansion of the Einstein-Hilbert action with a negative cosmological constant $\Lambda=-1$, i.e.,
\begin{equation}
S=\frac{1}{16\pi G}\int d^3x\sqrt{-g}(R-2\Lambda)
\end{equation} 
to the quadratic order around the background. Note that the zero order term is irrelevant to our purpose and the first order one vanishes due to the fact that the background metric arises as a solution to the above action. So what we need to know is merely the quadratic term, which is given by
\begin{eqnarray}
S_h&=&-\frac{1}{64\pi G}\int d^3x\sqrt{-g}h^{ab}(\nabla_a\nabla_c{h^c}_b+\nabla_b\nabla_c{h^c}_a-\nabla^c\nabla_ch_{ab}-\nabla_a\nabla_bh\nonumber\\
&&-2h_{ab}-g_{ab}\nabla^c\nabla^dh_{cd}+g_{ab}\nabla^c\nabla_ch),
\end{eqnarray}
whereby the equation of motion can be obtained by the variational principle as
\begin{equation}
\nabla_a\nabla_c{h^c}_b+\nabla_b\nabla_c{h^c}_a-\nabla^c\nabla_ch_{ab}-\nabla_a\nabla_bh
-2h_{ab}-g_{ab}\nabla^c\nabla^dh_{cd}+g_{ab}\nabla^c\nabla_ch=0
\end{equation}
with $h$ the trace of $h_{ab}$.
Note that the dynamics of graviton has the gauge freedom inherited from the diffeomorphism invariance of full theory. So with the transverse traceless gauge conditions $\nabla^ah_{ab}=0$ and $h=0$, the equation of motion can be reduced to
\begin{equation}
(\nabla^c\nabla_c+2)h_{ab}=0.
\end{equation}

Conventionally one uses the Faddeev-Popov trick to work out the determinant expression for one loop partition function. Here we would like to employ the alternative approach, which is proven to be more efficient\cite{GGV}. To achieve this, let us firstly decompose the metric as follows
\begin{equation}
h_{ab}=h^T_{ab}+\frac{1}{3}g_{ab}\alpha+\nabla_a\xi_b+\nabla_b\xi_a,
\end{equation}
where $h^T_{ab}$ satisfies the transverse traceless condition, and the last two terms come from  the diffeomorphism contribution\footnote{It is necessary to preserve the degrees of freedom when one performs such a decomposition as we do here, namely the degrees of freedom is six on both sides. Otherwise, one would run into a wrong determinant expression for one loop partition function by using such an alternative approach at the end of day. Put it another way, this sort of decomposition can be regarded as sort of change of variables, which is well defined only with the degrees of freedom preserved. Otherwise, the corresponding Jacobian would be degenerate somehow. }. With this decomposition, we have
\begin{eqnarray}
\int\mathcal{D}h_{ab}e^{S_h}&=&\int Z_{ghost}\mathcal{D}h^T_{ab}\mathcal{D}\alpha\mathcal{D}\xi_ae^{-\frac{1}{64\pi G}\int d^x\sqrt{-g}h^{Tab}(-\nabla^c\nabla_c-2)h^T_{ab}+\frac{2}{9}\alpha(\nabla^c\nabla_c-3)\alpha}\nonumber\\
&=&\int Z_{ghost}\mathcal{D}\xi_a[\mathrm{det}_2(-\nabla^c\nabla_c-2)]^{-\frac{1}{2}}[\mathrm{det}_0(\nabla^c\nabla_c-3)]^{-\frac{1}{2}},
\end{eqnarray}
which implies that the one loop partition function is given by throwing away the whole volume of diffeomorphism as
\begin{equation}
Z_{graviton}=Z_{ghost}[\mathrm{det}_2(-\nabla^c\nabla_c-2)]^{-\frac{1}{2}}[\mathrm{det}_0(\nabla^c\nabla_c-3)]^{-\frac{1}{2}},
\end{equation}
where $Z_{ghost}$ arises as the Jacobian corresponding to change of variables from $h_{ab}$ to $\{h^T_{ab},\alpha,\xi_a\}$.
Obviously, in order to nail down the one loop partition function for graviton, we are required to work out the specific expression for $Z_{ghost}$, which can be accomplished in Appendix B as
\begin{equation}
Z_{ghost}=[\mathrm{det}_1(-\nabla^a\nabla_a+2)]^{\frac{1}{2}}[\mathrm{det}_0(-\nabla^a\nabla_a+3)]^{\frac{1}{2}}.
\end{equation}
Thus the determinant expression for one loop partition function of graviton reads
\begin{equation}\label{graviton}
Z_{graviton}=[\mathrm{det}_2(-\nabla^c\nabla_c-2)]^{-\frac{1}{2}}[\mathrm{det}_1(-\nabla^a\nabla_a+2)]^{\frac{1}{2}}.
\end{equation}
\subsection{Gravitino}
Start from the gravitino action
\begin{equation}
S_\varphi=-\int d^3x\sqrt{-g}\bar{\varphi}_a(\gamma^{abc}\nabla_b\varphi_c+\hat{m}\gamma^{ab}\varphi_b)=-\int d^3x\sqrt{-g}\bar{\varphi}_a\epsilon^{abc}(\nabla_b\varphi_c-\hat{m}
\gamma_b\varphi_c).
\end{equation}
Here $\hat{m}^2=\frac{1}{4}$, $\bar{\varphi}=\varphi^\dagger\gamma^0$,
$\gamma^{ab}=\gamma^{[a}\gamma^{b]}$, and
$\gamma^{abc}=\gamma^{[a}\gamma^b\gamma^{c]}$ with $\gamma^a=e_I^a\gamma^I$, where $e_I^a$ form a set of orthogonal normal vector bases, and Gamma matrices are defined as $\gamma^{(I}\gamma^{J)}=\eta^{IJ}$ with the requirement $\gamma^0\gamma^{I\dagger}\gamma^0=\gamma^I$. Then the equation of motion for gravitino field can be obtained by the variational principle as
\ba
\gamma^{abc}\nabla_b\varphi_c+\hat{m}\gamma^{ab}\varphi_b=0.
\ea
Note that the above action is invariant under the gauge transformation $\delta\varphi_a=\nabla_a\zeta-\hat{m}\gamma_a\zeta$ as
\begin{eqnarray}
\epsilon^{abc}(\nabla_b\delta\varphi_c-\hat{m}\gamma_b\delta\varphi_c)&=&\epsilon^{abc}(\nabla_{[b}\nabla_{c]}\zeta+\hat{m}^2\gamma_{bc}\zeta)=\epsilon^{abc}(\frac{1}{8}R_{bcde}\gamma^{de}\zeta+\hat{m}^2\gamma_{bc}\zeta)\nonumber\\
&=&\epsilon^{abc}(-\frac{1}{4}+\hat{m}^2)\gamma_{bc}\zeta=0.
\end{eqnarray}
So we can always impose the covariant gauge condition on gravitino field as
$\gamma^a\varphi_a=\nabla^a\varphi_a=0$, which gives rise to
\ba
\gamma^{abc}\nabla_b\varphi_c=\gamma^a\gamma^b\gamma^c
\nabla_b\varphi_c+(-g^{ab}\gamma^c+g^{ca}\gamma^b-g^{bc}\gamma^a)\nabla_b\varphi_c=\gamma^b
\nabla_b\varphi^a, \ea
 and
 \ba
\hat{m}\gamma^{ab}\varphi_b=\hat{m}(\gamma^a\gamma^b-g^{ab})\varphi_b=-\hat{m}\varphi^a.
\ea
Thus the equation of motion can be simplified as \ba \label{seom}(\gamma^b
\nabla_b-\hat{m})\varphi_a=0.\ea

On the other hand, in order to obtain the determinant expression for one loop partition function of gravitino field, let us firstly decompose the gravitino field as
\begin{equation}
\varphi_a=\varphi^T_a+\frac{1}{3}\gamma_a\chi+\nabla_a\zeta-\hat{m}\gamma_a\zeta,
\end{equation}
where $\varphi^T_a$ satisfies the covariant gauge condition. With such a decomposition, we have
\begin{eqnarray}
\int \mathcal{D}\varphi_ae^{S_\varphi}&=&\int Z_{ghost}\mathcal{D}\varphi^T_a\mathcal{D}\chi\mathcal{D}\zeta e^{-\int d^3x\sqrt{-g}\bar{\varphi}^T_b(\gamma^a\nabla_a-\hat{m})\varphi^{Tb}-\frac{2}{9}\bar{\chi}(\gamma^a\nabla_a+3\hat{m})\chi}\nonumber\\
&=&\int Z_{ghost}d\zeta\mathrm{det}_\frac{3}{2}(\gamma^a\nabla_a-\hat{m})\mathrm{det}_\frac{1}{2}(\gamma^a\nabla_a+3\hat{m}),
\end{eqnarray}
where $Z_{ghost}$ is obtained in Appendix B as
\begin{equation}
Z_{ghost}=[\mathrm{det}_\frac{1}{2}(-\nabla^a\nabla_a+\frac{3}{4})]^{-1}.
\end{equation}
Thus the one loop partition function for gravitino field is given by
\begin{eqnarray}\label{gravitino}
Z_{gravitino}&=&\mathrm{det}_\frac{3}{2}(\gamma^a\nabla_a-\hat{m})\mathrm{det}_\frac{1}{2}(\gamma^a\nabla_a+3\hat{m})[\mathrm{det}_\frac{1}{2}(-\nabla^a\nabla_a+\frac{3}{4})]^{-1}\nonumber\\
&=&[\mathrm{det}_\frac{3}{2}(-\nabla^a\nabla_a-\frac{9}{4}]^\frac{1}{2}[\mathrm{det}_\frac{1}{2}(-\nabla^a\nabla_a+\frac{3}{4})]^{-\frac{1}{2} .
}\end{eqnarray}
Here we have used the following facts, i.e.,
\begin{equation}
\mathrm{det}(\gamma^a\nabla_a+\hat{m})=\mathrm{det}(\gamma^a\nabla_a-\hat{m}),
\end{equation} 
and
\begin{equation}
\mathrm{det}_s(\gamma^a\nabla_a+\hat{m})\mathrm{det}_s(\gamma^a\nabla_a-\hat{m})=\mathrm{det}_s[(\gamma^a\nabla_a+\hat{m})(\gamma^a\nabla_a-\hat{m})]=
\mathrm{det}_s(\nabla^a\nabla_a-\hat{m}^2+s+1),
\end{equation}
where the proof of the former can be found in the Appendix C, and the latter will be demonstrated in the next section.

\section{Normal modes for gravitino and other fields}
In this section, we will construct the spectrum of normal modes for gravitino field around our
AdS$_3$ in an algebraic way. In addition, for later computation, we also include the relevant result for other fields in the end.

Let us get started by acting on both sides of the equation of motion for gravitino field (\ref{seom}) with $\gamma^c\nabla_c+\hat{m}$, which gives rise to
\ba
0&=& (\gamma^c \nabla_c+\hat{m})(\gamma^b
\nabla_b-\hat{m})\varphi_a=(\gamma^c\gamma^b \nabla_c
\nabla_b-\hat{m}^2)\varphi_a \nonumber\\
&=&(g^{cb}\nabla_c \nabla_b+\gamma^{cb}\nabla_c
\nabla_b-\hat{m}^2)\varphi_a
=(\nabla^b\nabla_b+\gamma^{cb}\nabla_{[c}
\nabla_{b]}-\hat{m}^2)\varphi_a.\ea
Note that \ba \nabla_{[c}
\nabla_{b]}\varphi_a=\frac18R_{cbde}\gamma^{de}\varphi_a
+\frac12R_{cba}{}^d\varphi_d. \ea
Thus we can further have \ba \label{eoml}
0=(\nabla^b\nabla_b+\frac18R_{cbde}\gamma^{cb}\gamma^{de}-\hat{m}^2)\varphi_a
+\frac12\gamma^{cb}R_{cba}{}^d\varphi_d,
\ea
which can be reduced to
\begin{equation}
(\nabla^b\nabla_b+\frac{5}{2}-\hat{m}^2)\varphi_a=0
\end{equation}
on our AdS$_3$.

On the other hand, by the definition of Lie derivative acting on $\varphi_a$, i.e.,
\begin{equation}
\mathcal{L}_\xi\varphi_a=\xi^b\nabla_b\xi_a-\frac{1}{4}\gamma^{cd}\psi_a\nabla_d\xi_c+\varphi_b\nabla_a\xi^b,
\end{equation}
we have
 \ba \mathcal
{L}_X\mathcal {L}_Y\vp_f&=&X^a\nabla_a\mathcal
{L}_Y\vp_f-\frac14\gamma^{ab}\mathcal
{L}_Y\vp_f\nabla_bX_a+\mathcal {L}_Y\vp_a\nabla_f X^a\nn\\
&=&X^a\nabla_a(Y^c\nabla_c\vp_f-\frac14\gamma^{cd}\vp_f\nabla_dY_c+\vp_c\nabla_f Y^c)\nn\\
&&-\frac14\gamma^{ab}(Y^c\nabla_c\vp_f-\frac14\gamma^{cd}\vp_f\nabla_dY_c+\vp_c\nabla_f Y^c)\nabla_bX_a\nn\\
&&+(Y^c\nabla_c\vp_a-\frac14\gamma^{cd}\vp_a\nabla_dY_c+\vp_c\nabla_a Y^c)\nabla_\rho X^a \nn\\
&=&X^a(\nabla_aY^c)\nabla_c\vp_f+X^aY^c\nabla_a\nabla_c\vp_f
-\frac14\gamma^{cd}(\nabla_a\vp_f)X^a\nabla_dY_c-\frac14\gamma^{cd}\vp_fX^a\nabla_a\nabla_dY_c\nn\\
&&+(\nabla_a\vp_c)X^a\nabla_f Y^c+\vp_cX^a\nabla_a\nabla_f
Y^c\nn\\
&&-\frac14\gamma^{ab}(\nabla_c\vp_f)Y^c\nabla_bX_a+\frac1{16}\gamma^{ab}\gamma^{cd}\vp_f(\nabla_dY_c)\nabla_bX_a
-\frac14\gamma^{ab}\vp_c(\nabla_f Y^c)\nabla_bX_a\nn\\
&&+(\nabla_c\vp_a)Y^c\nabla_f
X^a-\frac14\gamma^{cd}\vp_a(\nabla_dY_c)\nabla_f
X^a+\vp_c(\nabla_a Y^c)\nabla_f X^a \nonumber\\
&=&X^a(\nabla_aY^c)\nabla_c\vp_f+X^aY^c\nabla_a\nabla_c\vp_f
-\frac14\gamma^{cd}(\nabla_a\vp_f)X^a\nabla_dY_c-\frac14\gamma^{cd}\vp_fX^a\nabla_a\nabla_dY_c\nn\\
&&+(\nabla_a\vp_c)X^a\nabla_f Y^c+\vp_cX^a\nabla_a\nabla_f
Y^c-\frac14\gamma^{ab}(\nabla_c\vp_f)Y^c\nabla_bX_a \nonumber\\
&&+\frac1{16}\gamma^{ab}\gamma^{cd}\vp_f[\nabla_d(Y_c\nabla_bX_a)-Y_c\nabla_d\nabla_bX_a]
-\frac14\gamma^{ab}\vp_c[\nabla_f (Y^c\nabla_bX_a)-Y^c\nabla_f\nabla_bX_a]\nn\\
&&+(\nabla_c\vp_a)Y^c\nabla_f
X^a-\frac14\gamma^{cd}\vp_a[\nabla_d(Y_c\nabla_f
X^a)-Y_c\nabla_d\nabla_fX^a] \nonumber\\
&&+\vp_c[\nabla_a( Y^c\nabla_f X^a)-Y^c\nabla_a\nabla_fX^a].
\ea
Whence it is not hard to show
\ba
\mathcal{L}^2\vp_f&=&Z^a{}_a{}^c\nabla_c\vp_f+H^{ac}\nabla_a\nabla_c\vp_f
-\frac14Z^a{}_d{}_c\gamma^{cd}\nabla_a\vp_f-\frac14\gamma^{cd}\vp_f R_{cdae}H^{ae}\nn\\
&&+Z^a{}_f{}^c\nabla_a\vp_c+\vp_cR^c{}_{fad}H^{ad}
-\frac14Z^c{}_{ba}\gamma^{ab}\nabla_c\vp_f \nonumber\\
&&+\frac1{16}\gamma^{ab}\gamma^{cd}\vp_f(\nabla_dZ_{cba}-R_{abde}H_c{}^e)
-\frac14\gamma^{ab}\vp_c(\nabla_f Z^c{}_{ba}-R_{abfd}H^{cd})\nn\\
&&+Z^c{}_f{}^a\nabla_c\vp_a-\frac14\gamma^{cd}\vp_a(\nabla_dZ_{cf}{}^a-R^a{}_{fde}H_c{}^e)
+\vp_c(\nabla_a Z^c{}_f{}^a-R^a{}_{fad}H^{cd}) \nn\\
&=&\frac14(\nabla^a\nabla_a\vp_f
-\frac14\epsilon^a{}_d{}_c\gamma^{cd}\nabla_a\vp_f+\epsilon^a{}_f{}^c\nabla_a\vp_c
-\frac14\epsilon^c{}_{ba}\gamma^{ab}\nabla_c\vp_f\nn\\
&&-\frac1{16}R_{abdc}\gamma^{ab}\gamma^{cd}\vp_f
+\frac14R_{abfd}\gamma^{ab}\vp^d+\epsilon^c{}_f{}^a\nabla_c\vp_a+\frac14R^a{}_{fdc}\gamma^{cd}\vp_a-R^a{}_{fad}\vp^d) \nn\\
&=&\frac14(\nabla^a\nabla_a\vp_f+\frac12\epsilon^a{}_c{}_d\gamma^{cd}\nabla_a\vp_f
-2\epsilon_f{}^{ac}\nabla_a\vp_c \nonumber\\
&&+\frac1{16}R_{abcd}\gamma^{ab}\gamma^{cd}\vp_f+\frac12R_{abfd}\gamma^{ab}\vp^d-R_{fd}\vp^d),
\ea
where we have used the identity $\nabla_a\nabla_b\xi_c=R_{cbad}\xi^d$ satisfied by any Killing vector field $\xi$.
Now substitute (\ref{eoml}) into the above equation, we can obtain
\begin{eqnarray}
\mathcal{L}^2\vp_f&=&\frac{1}{4}(\hat{m}^2\vp_f+\frac12\epsilon^a{}_c{}_d\gamma^{cd}\nabla_a\vp_f
-2\epsilon_f{}^{ac}\nabla_a\vp_c-\frac1{16}R_{abcd}\gamma^{ab}\gamma^{cd}\vp_f-R_{fd}\vp^d)\nonumber\\
&=&\frac{1}{4}(\hat{m}^2\vp_f-\gamma^a\nabla_a\vp_f
-2\gamma_f{}^{ac}\nabla_a\vp_c-\frac1{16}R_{abcd}\gamma^{ab}\gamma^{cd}\vp_f-R_{fd}\vp^d)\nonumber\\
&=&\frac{1}{4}(\hat{m}^2\vp_f-3\gamma^a\nabla_a\vp_f
-\frac1{16}R_{abcd}\gamma^{ab}\gamma^{cd}\vp_f-R_{fd}\vp^d)\nonumber\\
&=&\frac{1}{4}(\hat{m}^2\vp_f-3\hat{m}\vp_f
-\frac1{16}R_{abcd}\gamma^{ab}\gamma^{cd}\vp_f-R_{fd}\vp^d)\nonumber\\
&=&\frac{1}{4}(\hat{m}^2-3\hat{m}
+\frac{5}{4})\vp_f.\end{eqnarray}
Similarly, one can have
 \ba \mathcal {\bar{L}}^2\vp_f
&=&\frac14(\hat{m}^2+3\hat{m}+\frac54)\vp_f. \ea
Together with the fact that the Lie derivative via Killing vector fields commutes with the covariant derivative and Gamma matrices as well,  the above result implies that the space of solutions to the equation of motion for gravitino field forms one representation of SL(2,$R$) Lie algebra, which is characterized by the value of the Casimir. Actually the same situation occurs for other fields. Moreover, the whole result can be cast into a uniform pattern. Namely as to a field $\Phi$ with spin $s$ satisfying the Laplace like equation\footnote{Although it is explicitly proven to be true for gravitino field in this section and for other fields with spin no greater than $2$ in \cite{Zhang} for example, which turns out to be sufficient for our later calculation, we are believed that this uniform pattern is expected to be also true for fields with any other spin from both bulk and boundary points of view.}
\begin{equation}\label{laplace}
(\nabla^a\nabla_a-\hat{m}^2+s+1)\Phi=0,
\end{equation}
one can have
\begin{equation}
\mathcal{L}^2\Phi=\frac{\hat{m}^2-2s\hat{m}+s^2-1}{4}\Phi, \mathcal{\bar{L}}^2\Phi=\frac{\hat{m}^2+2s\hat{m}+s^2-1}{4}\Phi .
\end{equation}
Then we can construct the corresponding spectrum of normal modes in an algebraic way by starting from the highest weight mode which obeys the condition as follows
\begin{equation}
\mathcal{L}_{L_+1}\Phi^{(0,0)}=0, \mathcal{L}_{\bar{L}_+1}\Phi^{(0,0)}=0,\mathcal{L}_{L_0}\Phi^{(0,0)}_+=w_+\Phi^{(0,0)}, \mathcal{L}_{\bar{L}_0}\Phi^{(0,0)}=w_-\Phi^{(0,0)}
\end{equation}
where $\Phi^{(0,0)}$ can be always expressed as
\begin{equation}
\Phi^{(0)}=e^{-iw_+u+iw_-v}\Phi^{(0,0)}(\rho)
\end{equation}
with $\mathcal{L}_{L_0}\Phi^{(0,0)}(\rho)=\mathcal{L}_{\bar{L}_0}\Phi^{(0,0)}(\rho)=0$.  Then the resultant normal modes can be obtained as the infinite
 tower of descendant modes, i.e.,
 \begin{equation}
 \Phi^{(p,q)}=(\mathcal{L}_{\bar{L}_{-1}})^p(\mathcal{L}_{L_{-1}})^q\Phi^{(0,0)}
 \end{equation}
with $p, q=0, 1, 2, \cdot \cdot \cdot$. Note that $\bar{L}_0+L_0$ and $\bar{L}_0-L_0$ represent the energy and angular momentum respectively. So the 
 corresponding frequencies and angular momentum can be calculated out as
\begin{equation}
\omega^{(p,q)}=w_-+w_++p+q, j^{(p,q)}=w_--w_++p-q.
\end{equation}
Using the commutation relation (\ref{commutation}), one can show that the conformal weight is given by
\begin{equation}
w_+=\frac{1\pm(\hat{m}-s)}{2}, w_-=\frac{1\pm(\hat{m}+s)}{2}.
\end{equation}
As expected, the whole result is in good agreement with the prediction from the two dimensional CFT, where the conformal dimension  and spin of dual primary operator to the bulk field are given by\footnote{Note that the relevant result is symmetric with respect to $\hat{m}$ and $-\hat{m}$. So in what follows, we shall work solely on one of them, say the case in which the value of $\hat{m}$ is positive. On the other hand, to have the modes normalizable as it stands, both $w_+$ and $w_-$ are required to be non-negative, which is reasonable also from the boundary point of view, since otherwise the dual operator would possess the correlation functions increasing with the distance. }
\begin{equation}\label{dimension}
\Delta=w_++w_-=1\pm\hat{m}, s=|w_+-w_-|.
\end{equation}

\section{Determinant of Laplace like operator from normal modes and one loop partition function for $\mathcal{N}=1$ supergravity}
This section is to show how the determinant of Laplace like operator can be evaluated through the spectrum of normal modes we obtained in the last section and work out the one loop partition function for $\mathcal{N}=1$ supergravity as an application.

To proceed, let us recall the fact that the Euclidean determinant of Laplace like operator in Eq.(\ref{laplace}), i.e.,
\begin{equation}
D_s(\Delta)=\mathrm{det}_s(\nabla^a\nabla_a-\hat{m}^2+s+1)
\end{equation}
depends on the boundary condition, which is specified in terms of the conformal dimension $\Delta$ by the holographic recipe. Then as shown in \cite{DHS}, by continuing to complex
$\Delta$ and matching poles as well as zeros, one can determine $D_s(\Delta)$ through the spectrum of normal modes as follows
\begin{eqnarray}
D_s(\Delta)&=&e^{\mathrm{Pol}(\Delta)}\prod_{p,q}(2\sinh\frac{\omega^{(p,q)}}{2T})^{2n}=e^{\mathrm{Pol}(\Delta)}\prod_{p,q}(\frac
{1-e^{-\frac{\omega^{(p,q)}}{T}}}{e^{-\frac{\omega^{(p,q)}}{2T}}})^{2n}\nonumber\\
&=&e^{\mathrm{Pol}(\Delta)}\prod_{k\geq0}(\frac
{1-e^{-\frac{\Delta+k}{T}}}{e^{-\frac{\Delta+k}{2T}}})^{2n(k+1)}\end{eqnarray}
for bosonic fields and
\begin{eqnarray}
D_s(\Delta)&=&e^{\mathrm{Pol}(\Delta)}\prod_{p,q}(2\cosh\frac{\omega^{(p,q)}}{2T})^{2n}=e^{\mathrm{Pol}(\Delta)}\prod_{p,q}(\frac
{1+e^{-\frac{\omega^{(p,q)}}{T}}}{e^{-\frac{\omega^{(p,q)}}{2T}}})^{2n}\nonumber\\
&=&e^{\mathrm{Pol}(\Delta)}\prod_{k\geq0}(\frac
{1+e^{-\frac{\Delta+k}{T}}}{e^{-\frac{\Delta+k}{2T}}})^{2n(k+1)}\end{eqnarray}
for fermionic fields, where $T$ denotes the temperature for the thermal AdS$_3$,  $n$ is the number of degrees of freedom of integrated fields, and $\mathrm{Pol}(\Delta)$ represents the polynomial function of $\Delta$\footnote{The specific form of this polynomial function is irrelevant to our current purpose, though it can be fixed by taking the large $\Delta$ limit indeed\cite{DHS}.}. Furthermore, by absorbing various terms legitimately into $\mathrm{Pol}(\Delta)$, we end up with
\begin{equation}
D_s(\Delta)=e^{\mathrm{Pol}(\Delta)}\prod_{k\geq0}(1-q^{\Delta+k})^{2n(k+1)}
\end{equation}
for bosonic fields and 
\begin{equation}
D_s(\Delta)=e^{\mathrm{Pol}(\Delta)}\prod_{k\geq0}(1+q^{\Delta+k})^{2n(k+1)}
\end{equation}
for fermionic fields, where
\begin{equation}
q=e^{2\pi i\tau}
\end{equation}
with $\tau=\frac{i}{2\pi T}$.

Now it is time for us to evaluate the one loop determinant for $\mathcal{N}=1$ supergravity. To proceed, let us firstly read off the conformal dimension for spin $2$ and spin $1$ field from (\ref{graviton}) as
\begin{equation}
\Delta_2=2,\Delta_1=3,
\end{equation}
where (\ref{laplace}) and (\ref{dimension}) are used. Furthermore, by taking into account the fact that  the number of degrees of freedom is two for both integrated spin $2$ and spin $1$ fields, we have
\begin{eqnarray}\label{boson}
Z_{graviton}&=&\prod_{k\geq0}(1-q^{2+k})^{-2(k+1)}\prod_{k'\geq0}(1-q^{3+k'})^{2(k'+1)}\nonumber\\
&=&\prod_{k\geq0}(1-q^{2+k})^{-2(k+1)}\prod_{k'\geq1}(1-q^{2+k'})^{2k'}\nonumber\\
&=&\prod_{k\geq0}\frac{1}{(1-q^{2+k})^2},
\end{eqnarray}
where we have thrown away the irrelevant polynomial term $e^{\mathrm{Pol}(\Delta)}$.

Similarly, we can read off the conformal dimension for spin $\frac{3}{2}$ and spin $\frac{1}{2}$ field from (\ref{gravitino}) as
\begin{equation}
\Delta_\frac{3}{2}=\frac{3}{2}, \Delta_\frac{1}{2}=\frac{5}{2},
\end{equation}
whereby the one loop partition function for gravitino field is given by
\begin{eqnarray}\label{fermion}
Z_{gravitino}&=&\prod_{k\geq0}(1+q^{\frac{3}{2}+k})^{2(k+1)}\prod_{k'\geq0}(1+q^{\frac{5}{2}+k'})^{-2(k'+1)}\nonumber\\
&=&\prod_{k\geq0}(1+q^{\frac{3}{2}+k})^{2(k+1)}\prod_{k'\geq1}(1+q^{\frac{3}{2}+k'})^{-2k'}\nonumber\\
&=&\prod_{k\geq0}(1+q^{\frac{3}{2}+k})^2.
\end{eqnarray}
Combine (\ref{boson}) with (\ref{fermion}), we end up with the one loop partition function for $\mathcal{N}=1$ supergravity as
\begin{equation}
Z=\prod_{k\geq0}\frac{(1+q^{\frac{3}{2}+k})^2}{(1-q^{2+k})^2},
\end{equation}
which reproduces the result obtained by the heat kernel method in \cite{DGG}.

\section{Concluding remarks}
We have evaluated the one loop partition function for $\mathcal{N}=1$ supergravity in AdS$_3$ from scratch. In passing, we have also provided an explicit expression of one loop determinant for a field of arbitrary spin in AdS$_3$. Although the relevant result is in good agreement with the previous one in the literature, the method we have employed here is not only essentially new but also amazingly powerful, which makes the whole calculation as simple as possible. For one thing, instead of the conventional Faddeev-Popov trick, we work out the determinant expression for one loop partition function of $\mathcal{N}=1$ supergravity by the decomposition technique, which simplifies the involved calculation very much. For another, with the recently discovered  formula, the off shell one loop determinant can be remarkably expressed in terms of on shell quantities like normal modes, which turns out to be easy to construct in a purely algebraic way. In all, the whole procedure demonstrated here seems much simpler than the heat kernel method developed in \cite{DGG}, though there may be a close relation between them since the heat kernel is derived also by a group theoretical approach in \cite{DGG}.

We conclude with various generalizations of our work. First,  taking into account that the higher spin supergravity has recently been proposed in \cite{CHR}, one is tempted to exploit the procedure developed here to address the analogous issue for this theory in AdS$_3$. Second, when the black hole is present, the normal modes will be replaced by the quasinormal modes. As shown in \cite{Zhang}, the spectrum of quasinormal modes can also be constructed in a purely algebraic way, namely by the infinite tower of descendants of the chiral highest weight mode in the BTZ black hole. So it is highly interesting to see how the one loop determinant can be derived explicitly from quasinormal modes in the BTZ black hole, although it can also been obtained by a modular transformation from that in AdS$_3$ on general grounds\footnote{It is noteworthy that such an investigation has been recently made for higher spin fields in \cite{DD1} and \cite{DD2}, where nevertheless the relevant construction involves a little bit a posterori arguments.}. We hope to address these issues elsewhere in the near future.

\section*{Acknowledgements}
It is our pleasure to thank Thomas Creutzig, Matthias R. Gaberdiel, Daniel Grumiller, Rajesh Gopakumar,  Dmitri Vassilevich, and Xi Yin
for their helpful correspondences on various issues related to our project. HZ is also grateful to Elias Kiritsis, Rene Meyer, and Andy O'Bannon for their valuable discussions. HZ is supported by European Union grants FP7-REGPOT-2008-1-CreteHEP
 Cosmo-228644, and PERG07-GA-2010-268246 as well as EU program "Thalis" ESF/NSRF 
2007-2013. XZ is supported by NSFC (No.10975017) and the Fundamental Research
Funds for the Central Universities.

\section*{Appendices}

\section*{A. Some properties of Gamma matrices}
By definition of Gamma matrices, we have
\begin{eqnarray}
\gamma^a\gamma^b&=&g^{ab}+\gamma^{ab},\nonumber\\
\gamma^a\gamma^b\gamma^c&=&\gamma^{abc}+g^{ab}\gamma^c-g^{ca}\gamma^b+g^{bc}\gamma^a.
\end{eqnarray}
On the other hand, special to the three dimension, we have
\ba
\gamma^{ab}=\epsilon^{abc}\gamma_c,
\ea
whereby we can further have
\ba
\epsilon_{abd}\gamma^{ab}=\epsilon_{abd}\epsilon^{abc}\gamma_c
=-2\delta^c_d\gamma_c=-2\gamma_d,
\ea
and
\ba
\epsilon_{abd}\gamma^{ab}\gamma^d=\epsilon_{abd}\gamma^{abd}=-2\gamma_d\gamma^d=-6,
\ea
where the latter implies
\ba
\gamma^{abc}=\epsilon^{abc}.
\ea
In addition, we can also have
\begin{equation}
\gamma_{ab}\gamma^{ab}=\epsilon_{abc}\gamma^c\epsilon^{abd}\gamma_d=-2\delta^d_c\gamma^c\gamma_d=-2\gamma_d\gamma^d=-6.
\end{equation}

\section*{B. Explicit calculations for $Z_{ghost}$}

Let us firstly define the path integral measures for scalar, vector, and tensor fields by requiring
\begin{eqnarray}
\int \mathcal{D}\beta e^{-\langle\beta,\beta\rangle}&=&1,\nonumber\\
\int \mathcal{D}\mu_a e^{-\langle\mu,\mu\rangle}&=&1,\nonumber\\
\int \mathcal{D}\nu_{ab} e^{-\langle\nu,\nu\rangle}&=&1
\end{eqnarray}
with
\begin{eqnarray}
\langle\beta,\beta'\rangle&=&\int d^3x\sqrt{-g}\beta\beta',\nonumber\\
\langle\mu,\mu'\rangle&=&\int d^3x\sqrt{-g}\mu^a\mu'_a,\nonumber\\
\langle\nu,\nu'\rangle&=&\int d^3x\sqrt{-g}\nu^{ab}\nu'_{ab}.
\end{eqnarray}
Now decompose the vector field $\xi^a$ as
\begin{equation}
\xi^a=\xi^{Ta}+\nabla^a\lambda
\end{equation}
with $\nabla_a\xi^{Ta}=0$, then the corresponding Jacobian can be obtained as follows
\begin{eqnarray}\label{J1}
1&=&\int \mathcal{D}\xi_a e^{-\langle\xi,\xi\rangle}=\int J_1\mathcal{D}\xi^T_a\mathcal{D}\lambda e^{-\int d^3x\sqrt{-g}(\xi^{Ta}\xi^T_a-\lambda\nabla^a\nabla_a\lambda)}\nonumber\\
&=&J_1[\mathrm{det}_0(-\nabla^a\nabla_a)]^{-\frac{1}{2}},
\end{eqnarray}
where as we have done throughout our paper any total derivative term is always thrown away. On the other hand,  with the above decomposition, we have
\begin{equation}
h_{ab}=h^T_{ab}+\frac{1}{3}g_{ab}\alpha+\nabla_a\xi^T_b+\nabla_b\xi^T_a+2\nabla_a\nabla_b\lambda,
\end{equation}
which gives rise to
\begin{eqnarray}\label{J2}
1&=&\int\mathcal{D}h_{ab}e^{-\langle h,h\rangle}\nonumber\\
&=&\int J_2\mathcal{D}h^T_{ab}\mathcal{D}\alpha'\mathcal{D}\xi^T_a\mathcal{D}\lambda e^{-\int d^3x\sqrt{-g}[h^{Tab}h^T_{ab}+\frac{1}{3}\alpha'^2+2\xi^{Ta}(-\nabla^b\nabla_b+2)\xi^T_a+\frac{8}{3}\lambda(-\nabla^a\nabla_a+3)(-\nabla^b\nabla_b)\lambda]}\nonumber\\
&=&J_2[\mathrm{det}_1(-\nabla^a\nabla_a+2)]^{-\frac{1}{2}}[\mathrm{det}_0(-\nabla^a\nabla_a+3)]^{-\frac{1}{2}}[\mathrm{det}_0(-\nabla^a\nabla_a)]^{-\frac{1}{2}},
\end{eqnarray}
where $\alpha'=\alpha+2\nabla^a\nabla_a\lambda$. Combining (\ref{J1}) with (\ref{J2}) and taking into account $\mathcal{D}\alpha'=\mathcal{D}\alpha$ at the same time, we end up with
\begin{equation}
Z_{ghost}=\frac{J_2}{J_1}=[\mathrm{det}_1(-\nabla^a\nabla_a+2)]^{\frac{1}{2}}[\mathrm{det}_0(-\nabla^a\nabla_a+3)]^{\frac{1}{2}}.
\end{equation}

Similarly, let us define the path integral measures for fermion fields as
\begin{eqnarray}
\int\mathcal{D}\psi e^{-\langle\psi,\psi\rangle}&=&1,\nonumber\\
\int\mathcal{D}\varsigma_ae^{-\langle\varsigma,\varsigma\rangle}&=&1
\end{eqnarray}
with
\begin{eqnarray}
\langle\psi,\psi'\rangle&=&\int d^3x\sqrt{-g}\bar{\psi}\psi',\nonumber\\
\langle\varsigma,\varsigma'\rangle&=&\int d^3x\sqrt{-g}\bar{\varsigma}^a\varsigma'_a.
\end{eqnarray}
Then we have
\begin{eqnarray}
1&=&\int\mathcal{D}\varphi_ae^{-\langle\varphi,\varphi\rangle}=\int Z_{ghost}\mathcal{D}\varphi^T_a\mathcal{D}\chi\mathcal{D}\zeta e^{-\langle\varphi,\varphi\rangle}\nonumber\\
&=&\int Z_{ghost}\mathcal{D}\varphi^T_a\mathcal{D}\chi'\mathcal{D}\zeta e^{-\int d^3x\sqrt{-g}[\bar{\varphi}^T_a\varphi^{Ta}-\frac{1}{3}\bar{\chi}'\chi'+(\nabla_a\bar{\zeta}-\frac{1}{3}\nabla_b\bar{\zeta}\gamma^b\gamma_a)(\nabla^a\zeta-\frac{1}{3}\gamma^a\gamma^c\nabla_c\zeta)]}\nonumber\\
&=&\int Z_{ghost}\mathcal{D}\zeta e^{-\int d^3x \sqrt{-g}\bar{\zeta}(-\nabla_a+\frac{1}{3}\gamma^b\gamma_a\nabla_b)(\nabla^a-\frac{1}{3}\gamma^a\gamma^c\nabla_c)\zeta}\nonumber\\
&=&\int Z_{ghost}\mathcal{D}\zeta e^{-\frac{2}{3}\int d^3x\sqrt{-g}\bar{\zeta}(-\nabla^a\nabla_a+\frac{3}{4})\zeta}=Z_{ghost}\mathrm{det}_\frac{1}{2}(-\nabla^a\nabla_a+\frac{3}{4})
\end{eqnarray}
where  $\chi'=\chi-3\hat{m}\zeta+\gamma^a\nabla_a\zeta$. So we end up with
\begin{equation}
Z_{ghost}=[\mathrm{det}_\frac{1}{2}(-\nabla^a\nabla_a+\frac{3}{4})]^{-1}.
\end{equation}

\section*{C. Proof of $\mathrm{det}(\gamma^a\nabla_a+\hat{m})=\mathrm{det}(\gamma^a\nabla_a-\hat{m})$}
In order to prove the above identity, we are only required to show the  eigenvalues of $\gamma^a\nabla_a$ always come in pairs as $\{C,-C\}$, which can be demonstrated by acting on $\varphi_b$ in the following way. Namely assume $\gamma^a\nabla_a\varphi_b(t,\phi,\rho)=C\varphi_b(t,\phi,\rho)$ and $\varphi'_b(t,\phi,\rho)=\varphi_0(t,-\phi,\rho)(dt)_b-\varphi_1(t,-\phi,\rho)(d\phi)_b+\varphi_2(t,-\phi,\rho)(d\rho)_b=\varphi_b(t,\phi',\rho)|_{\phi'=-\phi}$, then we have
\begin{eqnarray}
\gamma^a\nabla_a\gamma^1\varphi'_b(t,\phi,\rho)&=&\gamma^a\gamma^1\partial_a\varphi'_c(t,\phi,\rho)\nonumber\\
&&+\frac{1}{4}\gamma^a\omega_{IJa}\gamma^{IJ}\gamma^1\varphi'_c(t,\phi,\rho)+\gamma^a\Gamma^c{}_{ab}\gamma^1\varphi'_c(t,\phi,\rho)\nonumber\\
&=&-\gamma^1\gamma^a\nabla_a\varphi_b(t,\phi',\rho)|_{\phi'=-\phi}=-C\gamma^1\varphi_b(t,\phi',\rho)|_{\phi'=-\phi}\nonumber\\
&=&-C\gamma^1\varphi'_b(t,\phi,\rho),
\end{eqnarray}
where we have used the fact that the non-vanishing Christofle symbols as well as non-vanishing spin connections are given by
\begin{eqnarray}
&&\Gamma^0{}_{02}=\frac{\sinh(\rho)}{\cosh(\rho)},\Gamma^1{}_{12}=\frac{\cosh(\rho)}{\sinh(\rho)}, \Gamma^2{}_{00}=\sinh(\rho)\cosh(\rho),\Gamma^2{}_{11}=-\sinh(\rho)\cosh(\rho),\nonumber\\
&&\omega_{02a}=-\sinh(\rho)(dt)_a,\omega_{12a}=\cosh(\rho)(d\phi)_a
\end{eqnarray}
with the choice of the orthogonal normal bases as $\{e_0^a=\frac{1}{\cosh(\rho)}(\frac{\partial}{\partial t})^a, e_1^a=\frac{1}{\sinh(\rho)}(\frac{\partial}{\partial \phi})^a, e_2^a=(\frac{\partial}{\partial\rho})^a\}$ in the coordinate system $\{t,\phi,\rho\}$.

So we have accomplished our proof by showing that $\gamma^1\varphi'_b$ is the eigenvector with eigenvalue $-C$ if $\varphi_b$ is the eigenvector with eigenvalue $C$\footnote{Actually this can be regarded as the analogous version of parity operation in AdS$_3$ background.}.

\section*{D. Some properties of Killing vector fields}
A Killing vector field $\xi^a$, by definition, is a vector field satisfying
\begin{equation}
\mathcal{L}_\xi g_{ab}=\nabla_a\xi_b+\nabla_b\xi_a=2\nabla_{(a}\xi_{b)}=0.
\end{equation}
Whence one can show such a Killing vector field also has the following nice properties, i.e.,
\begin{eqnarray}
\mathcal{L}_\xi\epsilon&=&0,\nonumber\\
\mathcal{L}_\xi\nabla_a&=&\nabla_a\mathcal{L}_\xi,\nonumber\\
\mathcal{L}_\xi\gamma^a&=&\gamma^a\mathcal{L}_\xi.
\end{eqnarray}
Now let us prove the first one by working on the three dimensional case as follows
\begin{equation}
\mathcal{L}_\xi\epsilon_{abc}=\xi^d\nabla_d\epsilon_{abc}+\epsilon_{dbc}\nabla_a\xi^d+\epsilon_{adc}\nabla_b\xi^d+\epsilon_{abd}\nabla_c\xi^d=\epsilon_{dbc}\nabla_a\xi^d+\epsilon_{adc}\nabla_b\xi^d+\epsilon_{abd}\nabla_c\xi^d.
\end{equation}
To achieve our proof, we only need to show that  the quantity $\epsilon^{abc}\mathcal{L}_\xi\epsilon_{abc}$ vanishes. This is the case because
\begin{equation}
\epsilon^{abc}(\epsilon_{dbc}\nabla_a\xi^d+\epsilon_{adc}\nabla_b\xi^d+\epsilon_{abd}\nabla_c\xi^d)=-6\nabla_d\xi^d=0.
\end{equation}
Next let us demonstrate why the Lie derivative via Killing vector fields commutes with both the covariant derivative and gamma matrices by acting on the spinor field in the following way, i.e.,
\begin{eqnarray}
\mathcal{L}_\xi\nabla_a\psi-\nabla_a\mathcal{L}_\xi\psi&=&\xi^b\nabla_b\nabla_a\psi+\nabla_b\psi\nabla_a\xi^b-\frac{1}{4}\gamma^{cd}\nabla_a\psi\nabla_d\xi_c-\nabla_a(\xi^b\nabla_b\psi-\frac{1}{4}\gamma^{cd}\psi\nabla_d\xi_c)\nonumber\\
&=&\xi^b(\nabla_b\nabla_a-\nabla_a\nabla_b)\psi+\frac{1}{4}\gamma^{cd}\psi\nabla_a\nabla_d\xi_c \nonumber\\
&=&\frac{1}{4}\xi^bR_{bacd}\gamma^{cd}\psi+\frac{1}{4}\gamma^{cd}\psi R_{cdab}\xi^b \nonumber\\
&=&\frac{1}{4}\xi^bR_{bacd}\gamma^{cd}\psi+\frac{1}{4} R_{abcd}\xi^b\gamma^{cd}\psi=0,
\end{eqnarray}
and
\begin{eqnarray}
\mathcal{L}_\xi\gamma^a\psi-\gamma^a\mathcal{L}_\xi\psi&=&\xi^b\nabla_b(\gamma^a\psi)-\gamma^b\psi\nabla_b\xi^a-\frac{1}{4}\gamma^{cd}\gamma^a\psi\nabla_d\xi_c-\gamma^a(\xi^b\nabla_b\psi
-\frac{1}{4}\gamma^{cd}\psi\nabla_d\xi_c) \nonumber\\
&=&-\gamma^b\psi\nabla_b\xi^a-\frac{1}{4}(\gamma^{cd}\gamma^a-\gamma^a\gamma^{cd})\psi\nabla_d\xi_c=-\gamma^b\psi\nabla_b\xi^a+g^{a[c}\gamma^{d]}\psi\nabla_d\xi_c \nonumber\\
&=&-\gamma^b\psi\nabla_b\xi^a+g^{ac}\gamma^d\psi\nabla_d\xi_c=0.
\end{eqnarray}


\begin{thebibliography}{99}

\bibitem{GMY}S. Giombi, A. Maloney and X. Yin,  {\it One-loop partition functions of 3D
gravity}, JHEP {\bf08} (2008) 007[arXiv:0804.1773].

\bibitem{DGG}J. R. David, M. R. Gaberdiel and R. Gopakumar, {\it The heat kernel on AdS$_3$ and its
applications},  JHEP {\bf04} (2010) 125[arXiv:0911.5085].
 
\bibitem{DHS}F. Denef, S. A. Hartnoll, and S. Sachdev, {\it Black hole determinants and quasinormal modes},
Class. Quant. Grav. {\bf27} (2010) 125001[arXiv:0908.2657].
 
\bibitem{MS}J. Maldacena and A. Strominger, {\it AdS$_3$ black holes and a stringy exclusion principle}, JHEP {\bf12} (1998) 005[arXiv:hep-th/9804085].

\bibitem{BKL}V. Balasubramanian, P. Kraus, and A. Lawrence, {\it Bulk vs. boundary dynamics in Anti-de Sitter spacetime}, Phys. Rev. D {\bf 59} (1999) 046003[arXiv:hep-th/9805171].

\bibitem{Zhang}H. Zhang, {\it SL(2,$R$) symmetry and quasi-normal modes in the BTZ black
hole}, JHEP {\bf03} (2011) 009[arXiv:1102.4721].

\bibitem{Wald}R. M. Wald, {\it General Relativity},  University of Chicago
Press, Chicago U.S.A. (1984).

\bibitem{GGV}M. R. Gaberdiel, D. Grumiller, D. Vassilevich, {\it Graviton 1-loop partition function for 3-dimensional massive gravity
},  JHEP {\bf11} (2010) 094[arXiv:1007.5189].

\bibitem{CHR}T. Creutzig, Y. Hikida and P. B.  Ronne, {\it Higher spin AdS$_3$ supergravity and its dual
CFT}, JHEP {\bf02} (2012) 109[arXiv:1111.2139].

\bibitem{DD1}S. Datta and J. R. David, {\it Higher spin quasinormal modes and one-loop determinants in the BTZ
black hole}, JHEP {\bf03} (2012) 079[arXiv:1112.4619].

\bibitem{DD2}S. Datta and J. R. David, {\it Higher spin fermions in the BTZ black hole}, arXiv: 1202.5831.



\end{thebibliography}
\end{document}